\documentclass[aps,pra,superscriptaddress,showpacs,noshowkeys]{revtex4}
\usepackage{epsfig}
\usepackage{graphicx}
\usepackage{bm}
\usepackage{amssymb}
\begin{document}
\newcommand{\Br}[1]{(\ref{#1})}
\newcommand{\Eq}[1]{Eq.\ (\ref{#1})}
\title{ The electric dipole moment of an electron in H-like ions in an electrostatic storage ring}%
\author{A. A. Bondarevskaya$^{1}$, D. V. Chubukov$^{1}$, O. Yu. Andreev$^1$, \\ E. A. Mistonova$^1$, L. N. Labzowsky$^{1, 2}$, G. Plunien$^3$, D. Liesen$^{4,5}$ and F. Bosch$^4$}
\affiliation{1 St. Petersburg
State University, Uljanovskaja 3,  198504, St. Petersburg, Petrodvorets,
Russia \\
2 Petersburg Nuclear Physics Institute, 188300, St. Petersburg,
Gatchina, Russia \\
3 Institut f\"{u}r Theoretische Physik, Technische Universit\"{a}t
Dresden, Mommsenstrasse 13, D-01062, Dresden, Germany \\
4 GSI Helmholzzentrum f\"{u}r Schwerionenforschung GmbH, Planckstrasse 1, D-64291,
Darmstadt, Germany \\
5 Fakult\"at f\"ur Physik und Astronomie,
Ruprecht-Karls-Universit\"at Heidelberg, D-69120, Heidelberg ,
Germany}
\begin{abstract}
Theoretical studies are presented how the electric dipole moment (EDM) of the electron in H-like ions
in electrostatic storage rings can sensitively be determined. With the proposed experiments a new constraint of about $10^{-29}$ e cm
for the electron EDM can be established what is by an order of magnitude more restrictive than the
existing bounds. Experiments with H-like ions may provide a possibility to distinguish between the electron EDM effect and the effect of P,T violating interaction between the atomic electron and the nucleus.
\end{abstract}


\maketitle
\section{Introduction}

The search for the electric dipole moments (EDMs) of particles (electrons, muons, neutrons,
protons) as well as the EDMs of closed many-particle systems (nuclei, atoms, molecules) presents
one of the most important fundamental physics problems starting from the works of Purcell and Ramsey
\cite{Purc50} on the neutron EDM and of Salpeter \cite{Salp58} on the electron EDM in atoms. The prediction
for the electron EDM following from the Standard Model (SM) is $d_e<10^{-38}$ e cm (e is the electron
charge) \cite{Posp91}. However, within various extensions of the SM \cite{Eng13} the electron EDM can be enhanced to the value close to the bounds obtained in the recent accurate experiments with heavy diatomic molecules YbF \cite{Huds11}, ThO \cite{ACME13}. A possibility for the observation of the charged particles EDMs in heavy diatomic molecules was first discussed in \cite{Sand67}, where the observation of the proton EDM in diatomic molecules with closed electron shells was considered. For the observation of the electron EDM the diatomic molecules with non-closed electron shells should be used. Due to the closely lying opposite parity $\Omega$-doublets in the heteronuclear diatomic molecules with one heavy nucleus and open electron shells the space parity violating (P-odd) \cite{Labz78} and space parity and time invariance violating (P, T-odd) effects, including the EDM effect \cite{Sush78}-\cite{Gorsh79} are strongly enhanced. For the extraction of a bound for the electron EDM from the molecular measurement complicated theoretical calculations are necessary. Such calculations for YbF, PbF, ThO and other molecules were performed in \cite{Dmit87}-\cite{Skripn13}.

Another idea to observe the muon EDM in magnetic storage rings was suggested in \cite{Bail78}. This idea consists of observation of the charged particle spin precession in an electric field due to the existence of the EDM. This idea was realized for muons in frames of the (g - 2) experiment with polarized muons in a magnetic storage ring \cite{Bail78} and led to a new constraint for the muon EDM. Several years ago it was suggested that this constraint can be essentially improved by the full compensation of the magnetic field in the rest frame of the particle by an external electric radial field \cite{Semer98} - \cite{Farley}. The same idea was discussed for bare nuclei and for highly charged
ions (HCI) with closed electron shells in \cite{Khripl98}, \cite{Khripl00}. Possible proton EDM experiments in magnetic storage rings were described in \cite{Semer08}. A proposal for the
observation of the electron EDM in H-like HCI in magnetic storage rings was made in \cite{BondPR11}.
Very recently it was suggested to use electrostatic storage rings for the observation of the charged particles
(muon, proton, deuteron) EDMs \cite{Semer09}. In this paper we extend this idea to the observation of the
electron EDM in the H-like ions.

The operating electrostatic storage rings, which however were never used for the search for EDMs, exist
in Aarhus (Denmark) \cite{Moll97}, Stockholm (Sweden) \cite{Ren04} and Heidelberg (Germany) \cite{Fad06}.
Very recently it was also suggested to employ an electrostatic storage ring to observe the electron EDM in molecular ions and the possibility to reach the boundary $10^{-30}$ e cm was anticipated \cite{Kaw11}. In \cite{Kaw11} an ion $WN^+$ was chosen for the theoretical studies. The proposed experiment in \cite{Kaw11} was very similar to the experimental techniques employed in \cite{Huds11},\cite{ACME13} with molecular beams but could provide better statistics. This experimental techniques differ essentially from the simple magnetic resonance \cite{Purc50} and the muon spin precession effect in the external electric field as in \cite{Bail78}. It is assumed that the electron spins are polarized by laser pumping to some excited state, then the electron spin rotates due to the EDM in the external electric field and this rotation is fixed in the decay process of the excited state. This scheme is quite similar to one suggested for the electron EDM observation in the H-like ions in magnetic storage rings in \cite{BondPR11}.

The electron EDM search with the H-like ions has important advantages compared to the electron EDM experiments with heavy molecules. First, the experiments with H-like ions require much simpler theoretical support. This will become quite important for the extraction of accurate values for the EDM from the experimental results after the experimental discovery of the electron EDM. Moreover, we will demonstrate that the experiments with H-like ions may provide a possibility to distinguish the electron EDM effect and the effect of P,T-odd interaction between the atomic electron and the nucleus. These two effects cannot be distinguished in experiments with heavy neutral atoms and molecules.

An electrostatic storage ring as described in \cite{Moll97} - \cite{Fad06} presents a ring with two deflection areas formed by pairs of plates (electrodes) as it is shown in Fig.1.  The electric field is sustained between these plates and has
a radial direction. This electric field compensate the centrifugal force and the injected charged particles move along a closed trajectory
in the ring. The radius of the ring R grows up with the mass and velocity of an ion but drops down with the larger ion charge. Assuming
the applied electric field of the order $\mathcal{E} \approx 10^5$ V/cm and the velocity of the particles of about
$0.1 c$ ($c$ is the speed of light) we obtain for the different ions the radii given in Table I. This Table demonstrates that
compared to the existing magnetic storage rings the electrostatic rings have the size smaller by an order
 of magnitude.

 \section{Polarization methods}

For the observation of the EDM effect in storage rings polarized particles are
necessary \cite{Bail78} - \cite{BondPR11}. Production, preservation and monitoring of polarized
H-like HCI beams in magnetic storage rings are discussed theoretically in \cite{BondPR11}. As a
production method a selective laser pumping of the excited hyperfine sublevel of a ground electronic
level of an ion was suggested in \cite{Proz03} (see also \cite{BondPR11}), where the $^{151}_{63}Eu^{62+}$
ion was  considered as an example. The selective laser pumping method of polarization consists of the excitation of a hyperfine sublevel of
H-like HCI with a circularly polarized optical laser. This excitation leads to an inhomogeneous occupation
of the Zeeman substates of the excited hyperfine state, i.e. the excited hyperfine level becomes polarized. This happens during one laser pulse ($\approx 5 \cdot 10^{-8}$ s), i.e. after one revolution of the ion around the ring ($\approx 10^{-6}$ s). The laser beam is assumed to travel parallel to the ion beam in a certain ring area. Then the quantization axis is called longitudinal and the laser produces a longitudinal polarization.
In \cite{Klaft94}, \cite{Seel98} resonant laser excitation measurements of the hyperfine structure of H-like $^{207}_{82}$Pb and $^{209}_{83}$Bi ions were performed. It follows from the results of these measurements that during one pulse an equilibrium between the excited and the ground hyperfine levels is established. After the ions leave the polarization area 50$\%$ of them will be in an excited (polarized) hyperfine state and 50$\%$ will be in the ground (unpolarized) hyperfine state of the ground electronic state. After the excited hyperfine state decays (see Table II for the decay time) the laser should be switched on again. A question arises, whether the ions will not lose their polarization in the process of decay? In \cite{BondPR11} it was proved that the ions will not lose the polarization in the particular case of transitions between the hyperfine levels $F \rightarrow F'$ when $F = F' + 1$. This case corresponds to all the transitions in Table II. In a more general way this problem is discussed in Appendix A. Calculations of the polarization dynamics show that initially unpolarized ions will acquire the 100$\%$ polarization after some time period $t_{pol}$ (see Table II). We define the degree of polarization $\lambda_F$ as \cite{BondPR11}
 \begin{equation}
 \label{1}
 \lambda_F = \frac{1}{F} \sum_{M_F} n_{FM_F} M_F,
 \end{equation}
 where $F$, $M_F$ are the quantum numbers for the total angular momentum and its projection for an ion and $n_{FM_F}$ are the occupation numbers for the Zeeman substates. It is assumed that
\begin{equation}
 \label{2}
\sum_{M_F} n_{FM_F} = 1.
\end{equation}
From Table II one can see that the selective laser pumping method should work only for relatively heavy H-like ions with
$Z \geq 30$, otherwise the 100$\%$ polarization time becomes too long and the necessary laser frequency is lying out of the optical region. The preservation of the ion polarization during the many revolutions of the ion around the ring is a hard problem due to the existence of the depolarization effects. Methods for the ion polarization preservation in the magnetic storage rings were discussed in \cite{BondPR11}. The corresponding methods for the electrostatic ring are essentially the same as for the preservation of the EDM effect and will be discussed below in section VI.

 \section{Electron spin precession in an electric field}

Creation of a longitudinally polarized H-like HCI beam in an electrostatic storage ring presents the first step of the proposed experiment (see Fig.1). The main aim of the EDM experiment is the generation of a
vertical component of the polarization due to the existence of the electron EDM. This is a second step of the experiment. For simplicity we start with the consideration of HCI with spinless nuclei. Then the H-like HCI will be treated as a particle with the mass $m = m_N A$, charge $q = (Z - 1)e$ ($A$ is the atomic number, $m_N$ is the nucleon mass, $Z$ is the nuclear charge number) and the magnetic moment equal to the magnetic moment of an electron $\vec{\mu} = \frac{2 \mu_B}{\hbar} \vec{s}$. Here $(-e)$ is the electron charge, $\vec{s}$ is the electron spin, $\hbar$ is the Planck constant, $\mu_B = \frac{e \hbar}{2m_ec}$ is the Bohr magneton and $m_e$ is the electron mass. If the particle possesses an EDM $\vec{d}$ which is directed along the particle spin, it undergoes a precession in an external electric field. This is the basic idea of all the EDM experiments in storage rings, magnetic or electrostatic. The EDM $\vec{d}$ and the spin of the particle $\vec{s}$ are connected via
\begin{equation}
 \label{3}
\vec{d} = \eta \frac{q}{2mc} \vec{s},
\end{equation}
where $\eta$ is a dimensionless constant which has to be determined in the experiment. In the rest frame of the ion
\begin{equation}
 \label{4}
\biggl( \frac{d\vec{s}}{dt} \biggr)_{\rm rest} = - \frac{\eta e}{2m_e c} \vec{s} \times [\vec{\mathcal{E}} + (\vec{\beta} \times \vec{\mathcal{H}}) -
\frac{\gamma}{\gamma + 1} \vec{\beta} (\vec{\beta} \cdot \vec{\mathcal{E}})] \equiv \vec{s} \times \vec{\Omega}_d ,
\end{equation}
where $\vec{\mathcal{E}}$, $\vec{\mathcal{H}}$ are the external electric and magnetic fields in the laboratory frame,
$\vec{\beta} = \vec{v}/c$, $\vec{v}$ is the particle velocity, $\gamma = 1/ \sqrt{1 - \beta^2}$. Spin precession occurs around the direction of the vector $\vec{\Omega}_d$, defined in Eq.(4).
In the nonrelativistic case $\vec{\Omega}_d$ coincides with $\vec{\mathcal{E}}$. Electron spin precession due to the electron magnetic moment is described by an equation (in the rest frame of an ion)
\begin{equation}
 \label{5}
\biggl( \frac{d\vec{s}}{dt} \biggr)_{\rm rest} = - \frac{e}{ m_e c} \vec{s} \times (\vec{\mathcal{H}} + \vec{\mathcal{H}}_m),
\end{equation}
where $\vec{\mathcal{H}}_m$ is the motional magnetic field
\begin{equation}
 \label{6}
\vec{\mathcal{H}}_m = \vec{\mathcal{H}}^{(1)}_m + \vec{\mathcal{H}}^{(2)}_m = - (\vec{\beta} \times \vec{\mathcal{E}}) - \frac{\gamma}{\gamma + 1} \vec{\beta} (\vec{\beta} \vec{\mathcal{H}}).
\end{equation}
This is actually the Bargmann-Michel-Telegdi (BMT) equation \cite{Barg59}, where the terms responsible for the Thomas precession are neglected. These terms depend on the particle acceleration and in case of H-like HCI are inversely proportional to the ion mass, i.e. negligible \cite{BondPR11}.

Equations (4), (5) can be generalized to any H-like HCI with a nucleus possessing the nuclear spin $\vec{I}$ \cite{BondPR11}. Then we have to replace in Eq.(3) the vector $\vec{s}$ by the vector $\vec{F} = \vec{I} + \vec{J}$, where $\vec{J}$ is the total angular momentum of the electron. Precession of the longitudinally polarized electron spin $\vec{s}$ around the direction of the radial
electric field due to the electron EDM leads to the generation of a vertical spin component (assuming that the ring is
horizontal, $x$ is radial direction, $y$ is longitudinal direction and $z$ is vertical direction). Due to the EDM
the ion polarization vector will rotate by a small angle $\varphi$ around the $x$ axis. This rotation will continue for every
revolution of the ions around the ring and the rotation angle $\varphi$ will grow up linearly with time (see Fig.2).

\section{The effect of electron EDM and the effect of P,T-violating interaction between atomic electron and nucleus }

The relation of the EDM of a neutral system (atom, molecule) to the EDMs of the charged particles (nuclei, electrons) incorporated in this system is regulated by the Schiff theorem \cite{Schiff63}. According to this theorem, the total electric field at the charged particle inside the neutral system is zero due to the electrostatic equilibrium, so that the EDM effect is absent. However, the Schiff theorem is violated either by strong interactions which determine the finite size of the nucleus, or by the magnetic interactions, i.e. relativistic corrections. As a result the nuclear EDM in atoms and molecules is suppressed compared to the bare nuclei, but the electron EDM in heavy atoms and molecules is, in contrary, enhanced due to the strong relativistic effects \cite{Sand65},\cite{Flam76}. Following \cite{Flam76}, we perform similar estimates for the HCI.

The "primary" EDM of an atom or ion due to the electron EDM is
\begin{equation}
 \label{7}
\vec{d_e}^{pr}=\langle 0 |\widehat{\vec{d_e}}^{pr}| 0 \rangle=d_e\langle 0 |\gamma_0 \vec{\Sigma}| 0 \rangle ,
 \end{equation}
where $\vec{\Sigma}$ are the Dirac matrices and the average $\langle 0| \ldots| 0 \rangle$ corresponds to the electronic state under consideration. The relativistic correction to the Stark matrix element
 \begin{equation}
 \label{8}
 S_{\rm EDM}=- \vec{d_e}^{pr} \vec{\mathcal{E}} ,
 \end{equation}
 where $\vec{\mathcal{E}}$ is an external electric field
 can be presented as \cite{Khrip97}, \cite{Khrip91}
 \begin{equation}
 \label{9}
 \delta S_{\rm EDM}=-d_e\langle 0 |\left( \gamma_0-1\right)\vec{\Sigma}\vec{\mathcal{E}}| 0 \rangle
 \end{equation}
 and estimated as
 \begin{equation}
 \label{10}
 |\delta S_{\rm EDM}|\approx d_e |\vec{\mathcal E} |\left(\alpha Z \right)^2 ,
 \end{equation}
 where  $\gamma_0$ is the Dirac matrix and $\alpha$ is the fine structure constant.
Due to the factor $(\gamma_0-1)$ only the lower component of the Dirac wave function contributes, i.e. the result is fully determined by relativistic effects.

 The electric dipole moment of an atom and accordingly the linear Stark matrix element
 \begin{equation}
 \label{11}
 S_{at}=e\langle 0 |\vec{r}\vec{\mathcal E} | 0 \rangle
 \end{equation}
are zero in the absence of the electron EDM. However, if the electron EDM is present, $S_{at}$ becomes nonzero due to mixing of the states with opposite parity by the interaction (in r.u.)
 \begin{equation}
 \label{12}
 \widehat{\vec{d_e}}^{pr}\vec{\mathcal E}_c=\widehat{\vec{d_e}}^{pr}\vec{r} \frac{Ze}{r^3} ,
 \end{equation}
where $\vec{\mathcal E}_c=\frac{Ze}{r^3}\vec{r}$ is the Coulomb field of the nucleus, $\vec{r}$ is radius-vector, $ r=|\vec{r}|$ and $\widehat{\vec{d_e}}^{pr}$ is defined in Eq.\Br{7}. Due to the interaction Eq.\Br{12} the atomic linear Stark effect becomes (we retain only relativistic corrections which give nonzero contribution to the EDM effect) :
\begin{eqnarray}
 \label{13}
\delta S_{at}&=&d_e e\vec{\mathcal E}\bigg[ \sum_n \frac{\langle 0 |\vec{r}|n \rangle\langle n |\left( \gamma_0-1\right)\vec{\Sigma}\vec{\mathcal E}_c |0\rangle }{E_n-E_0}+
\nonumber
\\
&&
+\sum_n \frac{\langle 0 |\left( \gamma_0-1\right)\vec{\Sigma}\vec{\mathcal E}_c |n\rangle\langle n |\vec{r}|0 \rangle }{E_n-E_0} \bigg] ,
 \end{eqnarray}
where $|n \rangle$, $E_n$ are the wave functions and energies of the atomic Dirac states with the parity opposite to the state $|0 \rangle$. In what follows we choose the state $|0 \rangle$ as the ground state $1s$ of the H-like ion. The sum over $n$ (only $p_{1/2}$ states are contributing) was evaluated exactly including the discrete and continuous Dirac spectra within the B-spline approach \cite{John88},\cite{Shab04}. For practical evaluations we consider the nucleus as a homogeneously charged sphere. The nuclear radii are taken from \cite{Ang04}. Using the order-of-magnitude estimates (in r.u.): $r\approx\frac{1}{m_e\alpha Z}$, $e|\vec{\mathcal{E}}_c|\approx m_e^2\left(Z\alpha\right)^3$, $(E_n-E_0)\approx m_e\left(\alpha Z\right)^2$, we obtain from Eq.\Br{13}
 \begin{equation}
 \label{15}
 |\delta S_{at}|\approx d_e |\vec{\mathcal{E}}|\left(\alpha Z\right)^2 ,
 \end{equation}
i.e. a correction of the same order as Eq.\Br{10}. Due to the Schiff theorem the nonrelativistic contributions to the total Stark shift Eq.\Br{8} and Eq.\Br{11} cancel out:
 \begin{equation}
 \label{16}
 S_{EDM}+S_{at}=0 .
 \end{equation}
Still the relativistic corrections Eq.\Br{9} and Eq.\Br{13} remain. The coefficient $\mathcal{K}_d$ for the effective electron EDM in an atom or ion can be defined as
 \begin{equation}
 \label{17}
 \mathcal{K}_d=\Bigg|\frac{\delta S_{EDM}+\delta S_{at}}{d_e |\vec{\mathcal{E}}|}\Bigg|.
 \end{equation}
It follows that for the case $|0\rangle=|1s\rangle$, $|n\rangle=|np\rangle$, the coefficient $\mathcal{K}_d\approx\left(\alpha Z\right)^2$, so  there is no enhancement. For another choice $|0\rangle=|2s\rangle$, $|n\rangle=|np\rangle$ using the estimate for the Lamb shift $(E_{np}-E_{2s})\approx m_e\alpha\left(\alpha Z\right)^4 $, we would have a strong enhancement, $\mathcal{K}_d\approx \frac{1}{\alpha\left(\alpha Z\right)^2}$. However, the EDM experiment with the excited $|2s\rangle$ state of the HCI as the basic one seems to be unrealistic, since the lifetime of the level $|2s\rangle$ in ions with high $Z$ value is quite short and is not enough to perform the EDM experiment (see Section V). The repeated excitation of the ions to the $|2s\rangle$ state should lead to the loss of polarization, i.e. also makes the EDM experiment impossible. The coefficients $\mathcal{K}_d$ for different ions are given in Table III.

There is an important advantage of the EDM experiment with different H-like HCI in storage rings compared to the linear Stark EDM measurements performed with certain neutral atoms and molecules. Collecting and comparing the results of the EDM experiments for the H-like ions with different $Z$ values one can distinguish the electron EDM effect from the effect caused by P- and T-violating interaction $V_{P,T}$ of the electron with the nucleus. This is not possible in experiments with neutral atoms or molecules. The identity of consequences of both effects was demonstrated in \cite{Gorsh79} for any particular atom or ion. However, the different dependence of these effects on $Z$ gives an opportunity to distinguish them provided that there is sufficient experimental data for the ions with different $Z$ values. The scalar P, T-violating interaction $V_{P,T}$ looks like \cite{Gorsh79},\cite{Khrip97},\cite{Khrip91}
 \begin{equation}
 \label{18}
 V_{ P,T}=Q_{P,T}g_{P,T}i\gamma_0 \gamma_5 \delta(\vec{r}),
 \end{equation}
where $\gamma_0, \gamma_5$ are Dirac matrices, $g_{P,T}$ is the constant of interaction and $Q_{P,T}$ is the "P,T-odd charge" of the nucleus. In the nonrelativistic limit operator $V_{P,T}$ takes the form (in r.u.)
\begin{equation}
 \label{18a}
V_{ P,T}=Q_{P,T}g_{P,T}i \frac{1}{2m_e}[\vec{\sigma} \hat{\vec{p}},\delta(\vec{r})],
\end{equation}
where $\vec{\sigma}$ are the Pauli matrices, $\hat{\vec{p}}$ is the momentum operator and
$[\,\, , \,\,]$ denotes the commutator.
The Stark shift caused by the interaction Eq.\Br{18} reads
 \begin{eqnarray}
 \label{19}
S_{P,T}&=&e\vec{\mathcal E}\bigg[ \sum_n \frac{\langle 0 |\vec{r}|n \rangle\langle n | V_{P,T}|0\rangle }{E_n-E_0}+
\nonumber
\\
&& +\sum_n \frac{\langle 0 |V_{P,T} |n\rangle\langle n |\vec{r}|0 \rangle }{E_n-E_0} \bigg].
 \end{eqnarray}
This expression we derived in the same way as Eq.\Br{13}.
The estimate for the electron momentum in an ion is $p\approx m_e\alpha Z$. For practical purposes we replace $\delta(\vec{r})$ in Eqs.\Br{18},\Br{19} by the nuclear density distribution $\rho_N(r)$. The expectation value $\langle\rho_N\rangle$ can be estimated as \cite{Khrip91} $\langle\rho_N\rangle\approx |\psi_{1s}(0)|^2R\approx\left(m_e \alpha Z\right)^3R$, where $\psi_{1s}(0)$ is the Schr\"{o}dinger wave function at the surface of the nucleus and $R$ is the relativistic enhancement factor. Then we obtain
 \begin{equation}
 \label{21}
S_{P,T}\approx Q_{P,T} m_e g_{P,T}e |\vec{\mathcal{E}}| \left(\alpha Z\right)R.
 \end{equation}
Here $R\approx \left(\frac{2 Z R_N}{a_0}\right)^{-\left(\alpha Z\right)^2}$ \cite{Khrip91}, $R_N$ is the nuclear radius and $a_0$ is the Bohr radius.

 Similar to Eq.\Br{17} we can define the coefficient $\mathcal{K}_{P,T}$ as
 \begin{equation}
 \label{21a}
\mathcal{K}_{P,T}=\Bigg|\frac{S_{P,T}}{m_e g_{P,T}e |\vec{\mathcal{E}}|}\Bigg|.
 \end{equation}
 Coefficients $\mathcal{K}_{P,T}(Z)$ for different H-like ions are also given in Table III.

For comparing the $Z$-dependence of the EDM effect and P,T-odd interaction effect we present the EDM effect in the form
\begin{equation}
 \label{23}
\xi_d=a_d \mathcal{K}_d\left(Z\right)  |\vec{\mathcal{E}}|,
 \end{equation}
where $\xi_d$ can be understood as the linear Stark shift or rotation angle in the external electric field $\vec{\mathcal{E}}$, $a_d$ is a numerical factor and $\mathcal{K}_d$ is a $Z$-dependent expression that can be exactly evaluated with Dirac wave functions. A similar expression for the P,T-odd interaction reads
 \begin{equation}
 \label{24}
\xi_{P,T}=a_{P,T} \mathcal{K}_{P,T}\left(Z\right) |\vec{\mathcal{E}}|.
 \end{equation}
 Then we can consider the ratios
  \begin{equation}
 \label{25}
R_d=\frac{\xi_d(Z=Z_1)}{\xi_d(Z=Z_2)}=\frac{\mathcal{K}_d(Z_1)}{\mathcal{K}_d(Z_2)},
 \end{equation}
 \begin{equation}
 \label{26}
R_{P,T}=\frac{\xi_{P,T}(Z=Z_1)}{\xi_{P,T}(Z=Z_2)}=\frac{\mathcal{K}_{P,T}(Z_1)}{\mathcal{K}_{P,T}(Z_2)}.
 \end{equation}
 For example, taking the simplest assumption $Q_{P,T}=A$ ($A=Z+N$, N is the number of neutrons, A is the atomic number) and comparing the $R$ values for $^{67}Zn^{29+}$ and $^{229}Th^{89+}$ we  have
 \begin{equation}
 \label{27}
R_d=\frac{\xi_d(Z=90)}{\xi_d(Z=30)}=13.3,
 \end{equation}
 \begin{equation}
 \label{28}
R_{P,T}=\frac{\xi_{P,T}(Z=90)}{\xi_{P,T}(Z=30)}=63.7.
 \end{equation}
 So the difference can be rather essential. The influence of the radiative corrections on the ratios \Br{27},\Br{28} should be negligible. Comparing the experimental $R$ value with the theoretical $R_d$ and $R_{P,T}$ values it should be possible, in principle, to distinguish between the electron EDM effect and the P,T-odd interaction effect.

 \section{Observation of the EDM effect}

In this paper we suggest the following scenario for the observation of the electron EDM. We suppose that the ions move in the ring within a time interval which is sufficient for the EDM rotation angle to grow up essentially.
 The angle $\varphi_{\rm EDM}$ after the observation time $t_{obs}$ becomes
\begin{equation}
\label{nonumber}
\varphi_{\rm EDM} = q |\vec{\omega}_d|\mathcal{K}_d t_{obs},
\end{equation}
where $|\vec{\omega}_d|$ is the frequency of the precession of the ion polarization around the vector $\vec{\Omega}_d$ (see Eq.(4)). This frequency can be estimated as \cite{Farley}
\begin{equation}
\label{29}
|\vec{\omega}_d| \approx e \eta |\vec{\mathcal{E}}|/2m_ec.
\end{equation}
A coefficient $q$ denotes the part of the ring where the ions move in the electric field. It is reasonable to assume $q\approx0.5$.
For an electric field of about $|\vec{\mathcal{E}}| \approx 10^5$ V/cm from Eq.\Br{29} follows  $|\vec{\omega}_d| = \eta\cdot 3\cdot10^{9} \, {\rm s}^{-1}$. For the
observation of the electron EDM at the level $10^{-29}$ e cm we should use
the value $\eta \approx 10^{-18}$ in our estimates. Then we can estimate $t_{obs}$ necessary to make
the rotation angle of the order of $10^{-4}\pi$ which seems to be a measurable value:
\begin{equation}
\label{30}
t_{obs} \approx 10^{-4} \pi(3\cdot 10^{9}q \mathcal{K}_d \eta )^{-1} \, s.
\end{equation}
The results obtained with formula (\ref{30}) are listed in Table III. After the EDM rotation angle has grown up sufficiently the third step of the experiment - the measurement of the effect should start. Polarization laser should be switched on again. For the ions with $Z\geq30$ the time $t_{obs}$ necessary for the observation of EDM is larger than the 100\% polarization time $t_{pol}$ and consequently much larger then the decay time $t_{dec}$. Then, at the start of the third step of the experiment all ions will be in their ground state. For the observation of the EDM effect it will be necessary to excite the ions back to the excited hyperfine sublevel. It can be done again with the same circularly polarized optical laser and this excitation will not destroy the existing polarization.

As we have already mentioned the decay of the excited hyperfine level with the total angular momentum value $F$ to the ground hyperfine level $F'$ will not destroy the ion polarization if $F>F'$ (see Appendix A). It remains to prove that the excitation process $F'\rightarrow F$ also will not destroy the ion polarization. Consider for example the case $F'=2$, $F=3$. Then for 100\% ion polarization only Zeeman substate with $M_{F'}=2$ will be occupied. A circularly polarized laser will be able to populate only $M_F=3$ Zeeman substate of the excited hyperfine level, i.e. 100\% polarization will be preserved.

We suggest to apply the same idea as for the HCI in a magnetic storage ring in \cite{BondPR11}: it will be necessary to measure the asymmetry in the number of the decay photons with fixed circular polarization with respect to the vertical polarization component of the ions. An expression for the transition probability of the decay process when the circular polarization of the emitted photons is fixed, looks like \cite{BondPR11}
\begin{equation}
\label{31}
dW = \frac{W_0}{4 \pi} [1 \pm \xi_{\rm EDM} Q (\vec{\zeta} \vec{\nu})],
\end{equation}
where $W_0$ is the total transition rate, $\vec{\zeta}$ is the unit vector of the ion beam polarization, $\vec{\nu}$ is the unit vector of the direction of the photon emission, $\pm$ correspond to the right (left) circular polarization of the emitted photons, $\xi_{\rm EDM}$ defines the magnitude of the EDM effect and $Q$ is a factor of order $1$, specific for the particular transition in an ion. The value of $\xi_{\rm EDM}$ can be presented as
\begin{equation}
\label{32}
\xi_{\rm EDM} = \lambda_F F \sin \varphi_{\rm EDM},
\end{equation}
where $F$ is the total angular momentum of the excited hyperfine sublevel of the ground electronic level of an ion, $\lambda_F$ is the degree of the ion polarization and $\varphi_{\rm EDM}$ is the EDM rotation angle in the xz plane.
It should be convenient to locate the photon detectors above and below the ring and to measure the asymmetry in the number of photons detected in the upper and lower hemispheres. The asymmetry $R$ equals to
\begin{equation}
\label{33}
R = 2 \xi_{\rm EDM} Q = 2 Q \lambda_F F \sin \varphi_{\rm EDM}.
\end{equation}
Knowing $R$ from the experimental data and using Eqs.\Br{32}, \Br{33} and \Br{29} one can define the $\eta$ (i.e. the electron EDM) value. In particular with $Q\left(1s_{1/2} F=3 \rightarrow 1s_{1/2} F=2\right)=\frac{1}{2}$ \cite{BondPR11},  $\varphi_{\mathrm{EDM}} \approx 10^{-4}\pi$ (see above), $\lambda_F = 1$ (since the time of observation is much greater than the polarization time, most of their time ions will move in the ring being fully polarized)   and $F$ values from the Table II, the $R$ values should be of the order $R \approx 10^{-3}$. For the observation of the effect of the order $10^{-3}$ over the level of fluctuations it is necessary to fix at least $10^6$ "events", i.e. HCI decays from the excited hyperfine $1s_{1/2}F$ level. For example, for $Z=63$ the decay time $t_{dec}=1.1\cdot 10^{-2}$s. Since this time is much smaller than the observation time $\approx 100$ hours for $Z=63$ ions, the necessary statistics for the fixation of the EDM effect in the electrostatic ring with $10^7$ ions in the beam can be obtained within 1 second even with the detector efficiency $10^{-3}$. For the registration of smaller asymmetry $R\approx 10^{-5}$ the observation time $t_{obs}$ for $Z=63$ ions will go down to $1$ hour. The registration time $t_{reg}$ then will grow up to $\approx 10^4$ s, i.e. will be approximately the same as $t_{obs}$.

 \section{Removal of the background effects}

The main advantage of the electrostatic ring compared to the magnetic one in the context of the electron EDM
experiments with H-like HCI is the reduction of the background effects. First, in the magnetic storage
ring the strong vertical bending magnets produce a fast electron spin rotation in the horizontal plane.
This leads to an averaging of the EDM effect to zero. Second, the focussing magnets which create a
magnetic field with radial component produce a false EDM effect. To overcome both difficulties a complicated
experimental scheme containing "Siberian snakes" was proposed in \cite{BondPR11}.

However in the rest frame of the ions a motional magnetic field arises in the electric field regions of the electrostatic ring. For the radial electric field the motional field will have a vertical component (see Eq.\Br{6}), i.e. will cause the same problems as the bending magnets in the magnetic
rings. In the electric field $10^5$ V/cm and for the ion velocity 0.1 c the motional field will be about $0.3\cdot 10^{-2}$ T (30 gs), i.e. essentially weaker than the magnetic fields of bending magnets in the magnetic storage rings (1 T). While passing the deflection area about few meters length with velocity 0.1 c the electron spin in H-like ion will rotate about $10$ times in the horizontal ($xy$) plane, i.e. the rotation angle will be about $20 \, \pi$. For the purpose of the EDM experiment we have to keep the ion polarization longitudinal, i.e. the deviation from the direction along the $y$ axis should not exceed $\pi/2$. We assume that the vertical component of the motional magnetic field $\vec{\mathcal{H}}_m^{(1)}$ can be compensated by real vertical magnetic field, i.e. to each pair of electrodes a vertical magnet can be attached that will compensate the influence of the motional field on the ion polarization. The insertion of such magnets in the electrostatic ring will partly diminish the strength of the electric
field acting on the moving ions, but this effect can be taken into account by the construction of the ring.

As it follows from Eq.(6) the motional field $\vec{\mathcal{H}}_m$ has two components: vertical $\vec{\mathcal{H}}^{(1)}_m$, presented by the first term in the right-hand side of Eq.(6), and longitudinal $\vec{\mathcal{H}}^{(2)}_m$, presented by the second term. Since the compensating magnetic field is orthogonal to the ion velocity, the second term in the right-hand side of Eq.(6) will be effectively suppressed. This term contains also a small parameter $\beta^2$.

To keep the ion polarization longitudinal after one revolution of the ions around the ring, the compensating magnetic field $\mathcal{H}_{c}$ should be fixed with an accuracy defined by the condition
\begin{equation}
\label{34}
\frac{\delta \mathcal{H}}{\mathcal{H}} = \frac{\mathcal{H}_c - \mathcal{H}}{\mathcal{H}} < 10^{-4}.
\end{equation}
For $\mathcal{H}_m \approx 30$ gs this gives $\delta \mathcal{H} \approx 3\cdot 10^{-3}$ gs which should be realistic. Possible systematic effects which grow up linearly with the number of revolutions could be suppressed by changing the direction of the ion velocity and simultaneous changing the ion polarization from longitudinal to anti-longitudinal one as is shown in Fig.3. The direction of the compensating magnetic field also should be reversed. If to perform such a change after every revolution the background effect from the vertical component of $\vec{\mathcal{H}}_m$ will be fully canceled but the EDM effect will survive: the vector $\vec{v}$ changes the sign, but $\vec{s}$ remains the same in Eqs.\Br{4}-\Br{6}. This would require a more complicated construction of the ring with switches that will change periodically the route of the bunched ion beam around the ring. The switches can be arranged by turning off few electrodes or turning on few additional ones. This will change the curvature of the trajectory of ions and hence change the route. The direction of polarization can be changed from longitudinal to anti-longitudinal and back with the spin rotator \--- a vertical magnet which rotates the electron spin in the horizontal plane by an angle $\pi$. If the spin rotators will be located outside the electrodes this operation will not do any harm to the EDM rotation angle.

The use of the cross-routes should strongly suppress the systematic background effects. The cancellation of the residual magnetic field (which remains after the compensation of the motional field by a real one) after two subsequent revolutions can be incomplete only due to the change of the velocity of the ion from one revolution to another. But this change $\frac{\delta v}{ v}$ is due to the fluctuations, i.e. grows up not linearly with the number of revolutions $N_{rev}$, but is proportional to $\sqrt{N_{rev}}$. Assuming that the field control in the ring can be as high as $10^{-4}$ we can expect that after the compensation of the motional magnetic field by the real one the residual magnetic field will be of the order $3\cdot 10^{-3}$ gs. Then after the cancellation of this residual field due to the cross-route scheme of the experiment the systematic background rotation angle will be of the order $3\cdot 10^{-3}\frac{\delta v}{ v}\pi$ per one revolution (the magnetic field of the 1 gs rotates the electron spin by an angle $\pi$ at the length interval of about few meters). Assuming that the velocity control is also of the order $\frac{\delta v}{ v}\approx10^{-4}$ we obtain a systematic "parasite" magnetic rotation $\delta \varphi_{syst}^{1rev}\approx 3\cdot 10^{-7}\pi$ per one revolution. Then with the $t_{obs}\approx 10^5$s ($Z=63$), i.e. for $10^{11}$ revolutions of the ions around the ring the total systematic background rotation angle will raise up to $\delta \varphi_{syst}^{tot}\approx 3\cdot 10^{-7}\cdot \sqrt{10^{11}}\pi\approx 0.1\pi$. This is acceptable since the vertical background magnetic field will not average out the EDM effect.

\section{Conclusions}

The feasibility of the proposed experiment depends mainly on the suppression of systematic errors which might be a difficult task. Still due to the relatively small size of the electrostatic ring it is possible to locate it into the cooler and into the vacuum chamber and to achieve full magnetic shielding. This could help to reach the required overall accuracy of about $10^{-4}$ for the field and ion velocity control.

A possible disadvantage of the electrostatic storage rings compared to the magnetic ones is the smaller ion current: maximum $10^7$ stored ions in the ring ELISA \cite{Moll97} compared to $10^{10}$ ions in ESR (GSI, Darmstadt). However, the number of ions (which determines the statistics of the experiment) is not so important for the proposed EDM experiment, since after the EDM rotation angle will reach its final value $\varphi_{\rm EDM} \approx 10^{-4}\pi$, the measurement of the photon emission asymmetry of the order $10^{-3}$ should not represent a serious difficulty. The smallness of the EDM effect in this experiment is reflected by the relatively large observation time $t_{obs}$ necessary to reach the rotation angle $\varphi_{\rm EDM} \approx 10^{-4}\pi$.

Another disadvantage of the electrostatic rings compared to the magnetic storage rings is the problem of the focusing. Due to the gravity the ions in the electrostatic ring will drop down essentially within few seconds. In $10^{-2}$ s the ion beam with $Z\approx 60$ will be shifted down by $1$ mm. To keep the beam at a certain height it would be necessary to switch on a "supporting" vertical electric field outside the deflection areas after some time period, for example after every $10^{-2}$ s, i.e. after $10^4$ revolutions of the ions around the ring. This supporting field can be rather weak: for the length of the non-deflecting area of about 1 m this field should be about 0.1 V/cm.

One more difficulty with the electrostatic rings is the relatively short lifetime of the beam in these rings. If in magnetic ring ESR the beam can exist for at least 10 hours, the lifetime of the beam in electrostatic ring is much shorter: 30 minutes in the ring DESIREE in Stockholm \cite{Thom11}. One of the reasons for this short lifetime is the loss of energy due to the radiation of the charged particle moving along the bended trajectory, i.e. electric bremsstrahlung in our case. This problem can be solved by the segments with longitudinal (accelerating) electric field in the ring. The longitudinal electric field should not produce any dangerous motional magnetic field provided that the field and velocity control conditions at the level $10^{-4}$ are satisfied.

\textbf{Acknowledgements}

A.B., O. A., E.M., L.L., G.P. and D.L. gratefully acknowledge the support by the DFG (grant SU 658/2-1).
A.B., O.A., E.M. and L.L. acknowledge the partial support by the RFBR grants 11-02-00168-a and 14-02-00188-a,  RFBR-DFG grant (RFBR 12-02-91340) and by the Ministry of Education and Science of Russian Federation, project 8420. The
work of A.B. and E.M. was supported by the FAIR-Russia Research Center Fellowship and by the nonprofit
foundation "Dynasty" (Moscow).
G.P. acknowledges the support from GSI Helmholzzentrum f$\ddot{\rm u}$r
Schwerionenforschung GmbH.

\setcounter{equation}{0}
\renewcommand{\theequation}%
{A.\arabic{equation}}

\section*{Appendix A: A conservation and loss of polarization in atomic transitions.}

Here we investigate the polarization behavior in atomic decay transitions $F \rightarrow F'$ where $F, F'$ are the total angular momenta of atomic states. For the final state $F'$ the degree of polarization defined by Eq.\Br{1} in the text reads
\begin{equation}
\label{b1}
\lambda_{F}' = \frac{1}{F'} \sum_{M_F'} M_F' n_{F'M_F'},
\end{equation}
where the occupation numbers can be presented via the transition probabilities $W_{FM_F \rightarrow F'M_F'}$ in the following way \cite{BondPR11}
\begin{equation}
\label{b2}
n_{F'M_F'} = \sum_{M_F} n_{FM_F} \frac{W_{FM_F \rightarrow F'M_F'}}{\Gamma_{FM_F}},
\end{equation}
$n_{FM_F}$ are the initial level occupation numbers and $\Gamma_{FM_F} = \sum_{M_F'} W_{FM_F \rightarrow F'M_F'}$ is the total width of the initial level $FM_F$.

Employing the Wigner-Eckart theorem for the matrix elements of transition probabilities
$$W_{FM_F \rightarrow F'M_F'} =$$
\begin{equation}
\label{b3}
 = \sum_{L M_L} C^{FM_F}_{F'M_F' \, LM_L} C^{FM_F}_{F'M_F' \, LM_L} |\langle F  \|  \textbf{V}_{L} \| F'  \rangle|^2,
\end{equation}
where $\langle F  \|  \textbf{V}_{L} \| F'  \rangle$ denotes the reduced matrix element of the photon-electron interaction operator $\textbf{V}_{L,M_{L}}$, $L, M_L$ being the emitted photon momentum and its projection, the degree of the final state polarization results as
$$\lambda_{F}' = \frac{1}{F'} \sum_{M_F} n_{FM_F} \times $$
\begin{equation}
\label{b4}
\times \sum_{M_F'LM_L} M_F' \frac{C^{FM_F}_{F'M_F' \, LM_L} C^{FM_F}_{F'M_F' \, LM_L} |\langle F  \| \textbf{V}_{L} \| F'  \rangle|^2}{\Gamma_{FM_F}}.
\end{equation}

Replacing the factor $M_F'$ in Eq.(A.4) by $M_F' = M_F - M_L$ we can perform the summation over the projection $M_F'$. For this purpose we rewrite the Clebsch-Gordan coefficients in terms of 3j-symbols and use the formula for summation over one angular momentum projection for two 3j-symbols \cite{Varsh88}
$$
\sum_{M_F'} (-1)^{F' - M_F'} \left (
\begin{array}{ccc}
F & L & F' \\
M_F & \overline{M}_L & \overline{M}_F'
\end{array}
\right ) \left (
\begin{array}{ccc}
F' & L & F \\
M_F' & M_L & \overline{M}_F
\end{array}
\right ) =
$$
\begin{equation}
\label{b4a}
= (-1)^{2F} \cdot \sum_{x} (-1)^{x} \Pi^2_x \left (
\begin{array}{ccc}
F & F & x \\
M_F & \overline{M}_F & 0
\end{array}
\right ) \left (
\begin{array}{ccc}
x & L & L \\
0 & M_L & \overline{M}_L
\end{array}
\right ) \times
\end{equation}
$$\times \left \{
\begin{array}{ccc}
L & L & x \\
F & F & F'
\end{array}
\right \} ,$$
where $\overline{m}_j=-m_j$, $\Pi_{a,b, \ldots c} = \sqrt{(2a + 1)(2b + 1)\ldots ((2c + 1))}$ and a standard notation for 6j-symbol is employed.

If only the dipole transition is under consideration ($L = 1$), due to the 6j-symbol properties $x = 0, 1, 2$.
It remains to perform the summation over $M_L$ and $x$ in two terms with factors $M_F$ and $M_L$.
In the first term only one 3j-symbol depends on $M_L$ and we can perform the summation of one 3j-symbol over one momentum projection \cite{Varsh88}:
\begin{equation}
\label{b4b}
\sum_{M_L} (-1)^{L - M_L} \left (
\begin{array}{ccc}
L & L & x \\
M_L & \overline{M}_L & 0
\end{array}
\right ) = \Pi_L \cdot \delta_{x 0}.
\end{equation}
Here $\delta_{xy}$ is the Kronecker delta.
What concerns the second term with the factor $M_L$ let's consider 3j-symbol with different $x$ parameter.
Here it is convenient to return to Clebsch-Gordan coefficients. For $x = 0$ the Clebsch-Gordan coefficient doesn't depend on the projection $M_L$:
\begin{equation}
\label{b4c}
C^{L \, M_L}_{L\, M_L  \, \, 00} = 1.
\end{equation}
For $x = 2$ the dependence is quadratic in $M_L$:
\begin{equation}
\label{b4d}
C^{L \, M_L}_{L\, M_L  \, \, 20} = \frac{3M_L^2 - L(L + 1)}{[(2L - 1)L(L + 1)(2L + 3)]^{1/2}}
\end{equation}
and only $x = 1$ gives the linear dependence on $M_L$:
\begin{equation}
\label{b4e}
C^{L \, M_L}_{L\, M_L  \, \, 10} = \frac{M_L}{L(L + 1)}.
\end{equation}
Due to the consequent multiplication by $M_L$ and summation over $M_L$ only the linear term survives.
This term ($x = 1$) gives linear dependence on $M_F$, so after collecting both terms (with factors $M_F$ and $M_L$) the degree of polarization of the final state can be presented in the following form:
\begin{equation}
\label{b5}
\lambda_{F}' = \lambda_{F} \cdot N(F, F'),
\end{equation}
where
\begin{equation}
\label{b6}
N(F, F') = \frac{F(F +1) + F'(F' + 1) - 2}{2F'(F + 1)},
\end{equation}
and $\lambda_{F}$ is the degree of polarization of the initial state.
In particular, choosing the momentum of the final state $F' = F - 1$ we have $N(F, F') = 1$, which means the conservation of the degree of polarization. For $F' = F$ and $F' = F + 1$ a coefficient $N(F, F') < 1$ would be obtained, which corresponds to the loss of polarization.

\begin{figure}[H]
\begin{center}
\includegraphics[width=8.5cm]{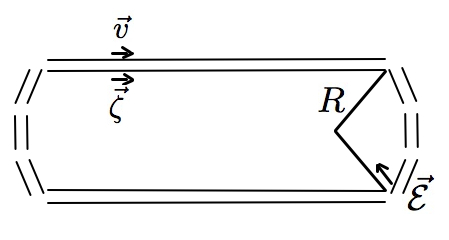}
\end{center}
\caption{\label{f:1} Electrostatic storage ring: $\vec{v}$ is the ion velocity, $\vec{\zeta}$ is the ion polarization (longitudinal), $\vec{\mathcal{E}}$ is the radial electric field, $R$ - radius of the ring, $||$ denotes the electrodes in the deflection area.}
\end{figure}

\begin{figure}[H]
\begin{center}
\includegraphics[width=8.5cm]{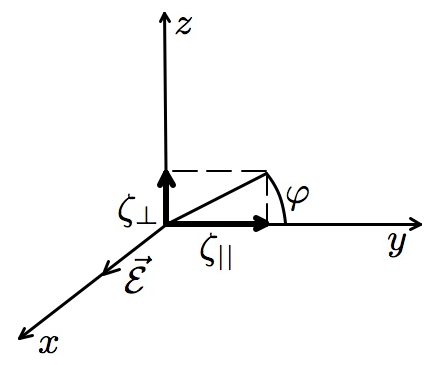}
\end{center}
\caption{\label{f:2} Electron spin rotation angle arising due to the existence of the EDM. $\zeta_{||}$ denotes the longitudinal (parallel to the ion velocity) component of the spin polarization vector, $\zeta_{\bot}$ denotes the vertical component arising due to the electron EDM. Axes $x, \, y, \, z$ are oriented along the radial, longitudinal and vertical directions, respectively.}
\end{figure}

\begin{figure}[H]
\begin{center}
\includegraphics[width=8.5cm]{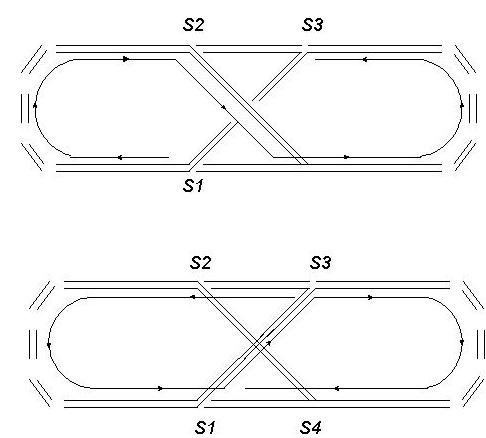}
\end{center}
\caption{\label{f:6} Schemes showing the routes of the bunched electron beam during the 1st and 2nd revolutions around the ring. $S_i$ ($i=1,2,3,4$) denote the switches. Upper panel: the first revolution, the ions start from the switch 1 and end up at the switch 3. Lower panel: the second revolution, the ions start from the switch 3 and end up at the switch 1.}
\end{figure}
\begin{table} [H]
\caption{Radii of the electrostatic storage rings $R$  for the different ions. The electric field strength is assumed to be
$\mathcal{E} \approx 10^5$ V/cm, the ion velocity 0.1 c (c is the speed of light).}
\begin{tabular}{cccc}
\hline\hline $Z$ &Ion & $R(m)$ \\
\hline 3 &  $^{7}{\rm Li}^{2+}$ & 3.29  \\
\hline 11 & $^{23}{\rm Na}^{10+}$ & 2.16  \\
\hline 19 & $^{39}{\rm K}^{18+}$ & 2.03 \\
\hline 30 & $^{67}{\rm Zn}^{29+}$ & 2.17  \\
\hline 41 & $^{93}{\rm Nb}^{40+}$ & 2.18 \\
\hline 51 & $^{121}{\rm Sb}^{50+}$ & 2.27  \\
\hline 63 & $^{151}{\rm Eu}^{62+}$ & 2.29  \\
\hline 71 & $^{175}{\rm Lu}^{70+}$ & 2.35 \\
\hline 79 & $^{197}{\rm Au}^{78+}$ & 2.37  \\
\hline 90 & $^{229}{\rm Th}^{89+}$ & 2.42 \\
\hline
\end{tabular}
\label{table:1}
\end{table}
\begin{table} [H]
\caption{Polarization characteristics of H-like HCI with the selective laser pumping polarization method} \centering
\footnotesize
\begin{tabular}{cccccc}
\hline\hline Nuclear & Isotope & Nuclear & Hyperfine
 & Decay time of the excited & 100\%
\\
charge & \, & spin & transition ($F \rightarrow F'$) & HF level ($F \rightarrow F'$ transition) &
polarization time
\\
Z & \, & I & frequency (eV)  & $t_{dec}$ (s) & $t_{pol}$(s)\\
\hline 3 & $^{7}{\rm Li}^{2+}$ & $3/2$ & $(2\rightarrow 1)$  $1.24 \cdot 10^{-4}$  & $1.4\cdot 10^{10}$ & $4.69 \cdot 10^{11}$ \\
\hline 11 & $^{23}{\rm Na}^{10+}$ & $3/2$ & $(2\rightarrow 1)$  $4.18 \cdot 10^{-3}$  & $3.67\cdot 10^5$ & $1.21 \cdot 10^{7}$ \\
\hline 19 & $^{39}{\rm K}^{18+}$ & $3/2$ & $(2\rightarrow 1)$ $3.88 \cdot 10^{-3}$  & $4.59\cdot 10^5$ & $1.51 \cdot 10^{7}$ \\
\hline 30 & $^{67}{\rm Zn}^{29+}$ & $5/2$ & $(3\rightarrow 2)$ $3.22 \cdot 10^{-2}$  & $1.12\cdot 10^3$ & $4.24 \cdot 10^{4}$ \\
\hline 41 & $^{93}{\rm Nb}^{40+}$ & $9/2$ & $(5\rightarrow 4)$ $5.74 \cdot 10^{-1}$  & $3.12\cdot 10^{-1}$ & $1.21 \cdot 10^1$ \\
\hline 51 & $^{121}{\rm Sb}^{50+}$ & $5/2$ & $(3\rightarrow 2)$  $0.71$  & $1.05\cdot 10^{-1}$ & $3.96$ \\
\hline 63 & $^{151}{\rm Eu}^{62+}$ & $5/2$ &  $(3\rightarrow 2)$ $1.51$  & $1.09\cdot 10^{-2}$ & $4.36 \cdot 10^{-1}$ \\
\hline 71 & $^{175}{\rm Lu}^{70+}$ & $7/2$ & $(4\rightarrow 3)$ $1.58$  & $1.22\cdot 10^{-2}$ &  $4.9 \cdot 10^{-1}$ \\
\hline 79 & $^{197}{\rm Au}^{78+}$ & $3/2$ & $(2\rightarrow 1)$ $0.19$ & $3.91$ & $1.2 \cdot 10^{2}$ \\
\hline 90 & $^{229}{\rm Th}^{89+}$ & $5/2$ & $(3\rightarrow 2)$ $1.11$  & $2.74\cdot 10^{-2}$ & $1.04$ \\\hline
\end{tabular}
\label{table:2}
\end{table}

\begin{table} [H]
\caption{The coefficients $\mathcal{K}_d$, $\mathcal{K}_{P,T}$ and the observation time for the electron EDM experiment in electrostatic ring for different H-like ions. The value for $t_{obs}$ is given
for the  asymmetry $R \approx 10^{-3}$, assuming that the electron EDM value is $\approx 10^{-29}$ e cm} \centering
\begin{tabular}{ccccc}
\hline\hline Nuclear & Ion & $\mathcal{K}_d$ &$\mathcal{K}_{P,T}$ & $t_{obs}$ \\
charge & \, &
\,& (in $Q_{P,T}$ units )& (s) \\
Z & \, &
\, \\
\hline 3 & $^{7}{\rm Li}^{2+}$ & 0.0011& $2.84\cdot 10^{-5}$ & $1.9\cdot 10^8$ ($\approx$ 6 years)  \\
\hline 11 & $^{23}{\rm Na}^{10+}$ &0.015 & $1.03\cdot 10^{-4} $&  $1.4\cdot 10^7$  ($\approx$ 161 days)  \\
\hline 19 & $^{39}{\rm K}^{18+}$ &0.045 & $2.09\cdot 10^{-4} $& $4.6\cdot 10^6$ ($\approx$ 54 days)  \\
\hline 30 & $^{67}{\rm Zn}^{29+}$ &0.12 &  $4.14\cdot 10^{-4} $& $1.8\cdot 10^6$ ($\approx$ 21 days)  \\
\hline 41 & $^{93}{\rm Nb}^{40+}$ &0.22 &   $7.22\cdot 10^{-4} $& $9.5\cdot 10^5$ ($\approx$ 263 hours)  \\
\hline 51 & $^{121}{\rm Sb}^{50+}$ &0.36 & $1.16\cdot 10^{-3} $ & $5.9\cdot 10^5$ ($\approx$ 163 hours) \\
\hline 63 & $^{151}{\rm Eu}^{62+}$ &0.58 & $2.04\cdot 10^{-3} $ & $3.6\cdot 10^5$ ($\approx$ 100 hours) \\
\hline 71 & $^{175}{\rm Lu}^{70+}$ &0.78 &  $2.99\cdot 10^{-3} $ & $2.7\cdot 10^5$ ($\approx$ 74 hours) \\
\hline 79 & $^{197}{\rm Au}^{78+}$ &1.04 &  $4.41\cdot 10^{-3} $ & $2.0\cdot 10^5$ ($\approx$ 56 hours) \\
\hline 90 & $^{229}{\rm Th}^{89+}$ &1.53 &  $7.71 \cdot 10^{-3} $ & $1.4\cdot 10^5$ ($\approx$ 38 hours) \\
\hline
\end{tabular}
\label{table:3}
\end{table}
\end{document}